\documentclass[aip,jap,reprint]{revtex4-1}

\usepackage{graphicx}
\usepackage{dcolumn}
\usepackage{bm}

\usepackage[utf8]{inputenc}
\usepackage[T1]{fontenc}
\usepackage{mathptmx}
\usepackage{etoolbox}
\usepackage{xcolor}

\begin{document}


\title{Gate-Drain Leakage Enhanced by Drain-Induced Dielectric Barrier Lowering in Gate-All-Around Field Effect Transistors}

\author{Juan P. Mendez}
 \email[]{jpmende@sandia.gov}
 \affiliation{Sandia National Laboratories, New Mexico 87123, USA}

\author{Coleman Cariker}
 \email[]{cbcarik@sandia.gov}
 \affiliation{Sandia National Laboratories, New Mexico 87123, USA}

 \author{Michael Titze}
 \affiliation{Sandia National Laboratories, New Mexico 87123, USA}
 \affiliation{Luxembourg Institute of Science and Technology (LIST), Belvaux 4422, Luxembourg}

 \author{Alex A. Belianinov}
\affiliation{Sandia National Laboratories, New Mexico 87123, USA}

\author{Denis Mamaluy}
 \email[]{mamaluy@sandia.gov}
\affiliation{Sandia National Laboratories, New Mexico 87123, USA}

\date{\today}

\begin{abstract}
Gate-All-Around Field-Effect Transistors (GAAFETs), now entering high-volume production as successors to fin field-effect transistor technology, are enabling continued scaling and enhanced performance in advanced semiconductor nodes. However, the drain-current in GAAFETs strongly deviates from the thermionic dependence at negative gate voltages, exhibiting the existence of leakage that is additionally enhanced at high applied biases. Understanding the origin of this leakage is essential for determining the scaling limits of GAAFETs and for guiding device and material optimizations aimed at suppressing the off-state current. Additionally, recent experimental measurements have revealed the increased influence of radiation-induced defects in the negative gate voltage regime, with their impact remaining largely negligible for positive gate voltages. Through predictive first-principles simulations, we demonstrate that the observed leakage current at negative gate voltages originates from gate-to-drain tunneling, which is significantly enhanced by drain-induced dielectric barrier lowering between the gate and drain.
\end{abstract}

\maketitle

\section{Introduction}\label{sec:introduction}

\begin{figure}[h]
\includegraphics[width=\columnwidth]{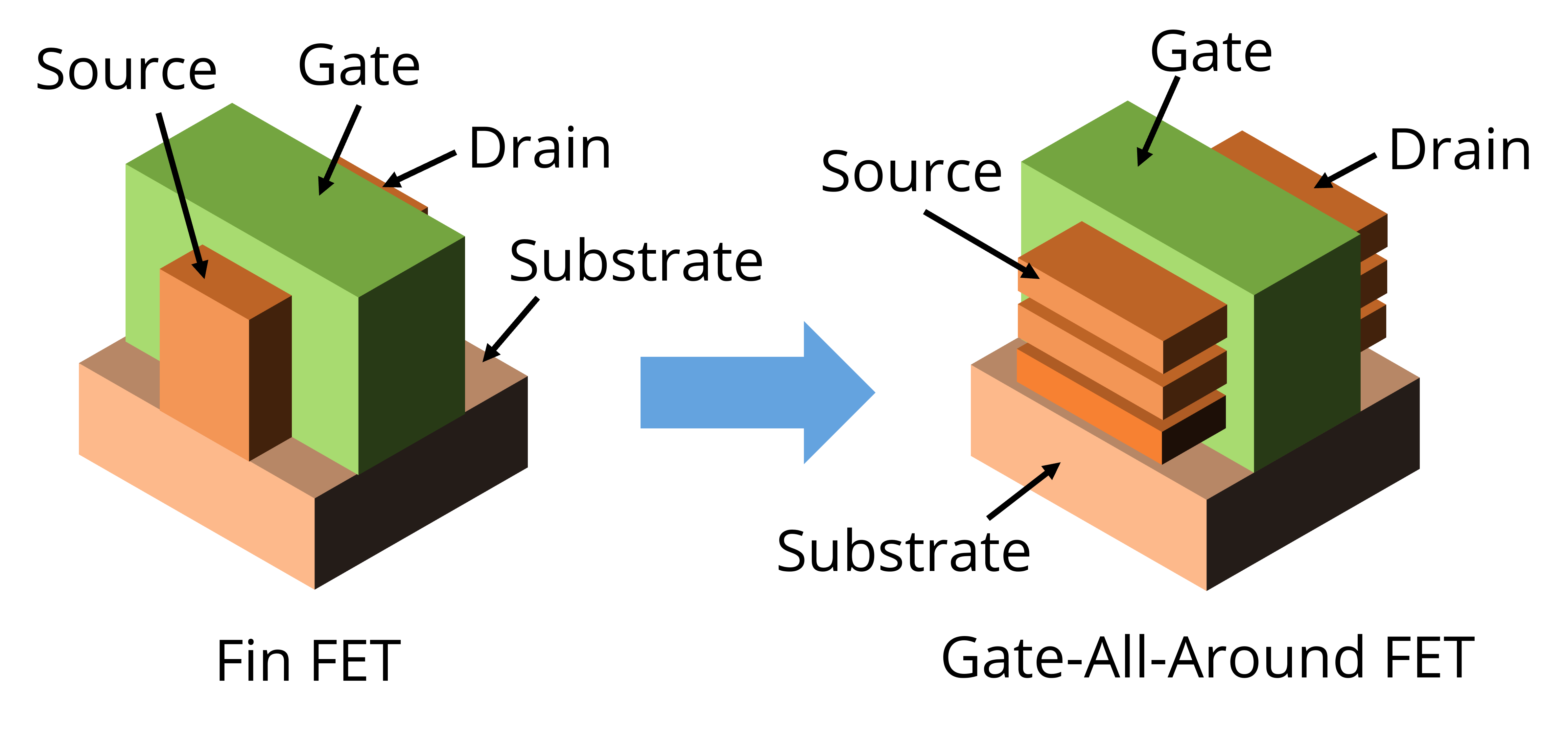}
\caption{Schematic representation of the technology transition from FinFET to next-generation GAAFET architecture.}
\label{Fig: FinFET_and_GAAFET}
\end{figure}

The semiconductor industry is undergoing a major transition from the mature Fin Field-Effect Transistor (FinFET) architecture to the next-generation Gate-All-Around Field-Effect Transistor (GAAFET) (Fig.~\ref{Fig: FinFET_and_GAAFET}), marking a pivotal milestone in the evolution of advanced computing systems \cite{IEEE_International_Roadmap_for_Devices_and_Systems2023}. This transition is already underway for several 3~nm technology nodes and is expected to be fully adopted across all 2~nm nodes \cite{CMOS_2nm}. Unlike FinFETs, where the channel is formed on a fin connected to the substrate, GAAFETs feature a fully isolated nanosheet or nanowire channel that are completely surrounded by the gate, offering superior electrostatic control, reduced leakage, and enhanced scalability for continued device miniaturization. This architecture also enables the realization of stacked channel structures, providing a pathway toward higher drive currents and increased integration density\cite{CMOS_2nm}.

\begin{figure}[t!]
\includegraphics[width=\columnwidth]{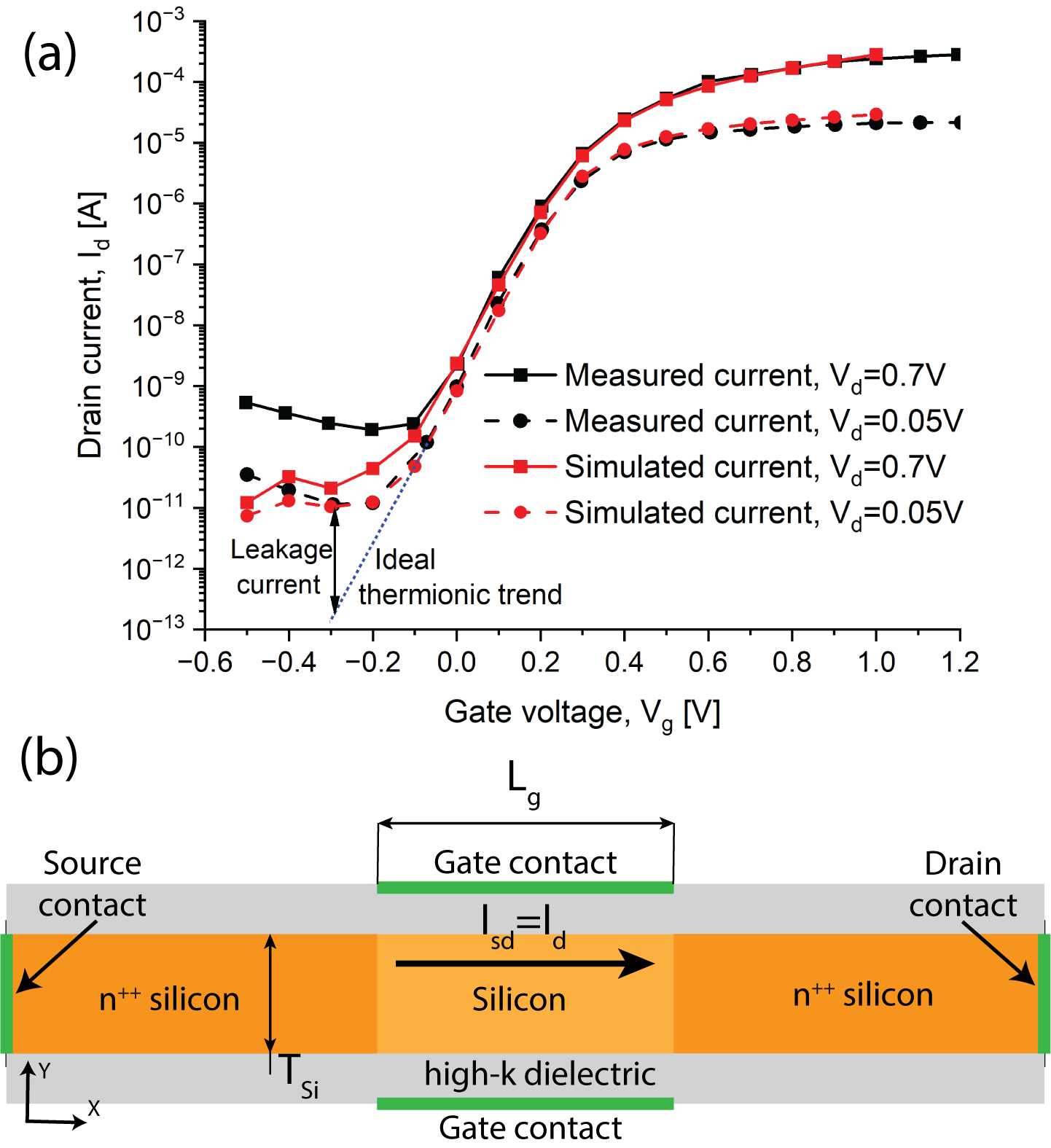}
\caption{(a) Measured drain-current/gate-voltage characteristics in Ref.~\cite{Loubet:2017} for a 44~nm-CPP GAAFET at two drain biases of $V_d=0.05$ and $0.7$~V, corresponding to the linear and saturation region operations, respectively, together with the simulated characteristics for an ideal-operation GAAFET shown in (b). Measured current reprinted with permission from Ref.~\citenum{Loubet:2017}. (b) Cross-section schematic of the simulated single-sheet GAAFET used for the ideal operation, in which gate leakage is negligible. $T_{Si}$ represents the thickness of the silicon channel.}
\label{Fig: ideal GAAFET}
\end{figure}

It was first demonstrated in 2017 that GAAFETs are a good candidate for the replacement of FinFETs \cite{Loubet:2017}.  Fig.~\ref{Fig: ideal GAAFET} includes the device drain-current/gate-voltage characteristics for a 44~nm-CPP (Contacted Poly Pitch) GAAFET, fabricated by IBM and reported in Ref.~\cite{Loubet:2017}, for two drain biases, corresponding to the linear and saturation region operations. Other companies, such as Intel, TSMC and Samsung are competing to bring this technology to mass production and commercial markets. As can be seen from Fig.~\ref{Fig: ideal GAAFET}, in the operating regime (for gate voltages in the range of $V_{g}=0$-$0.7$~V), the device exhibits the characteristic of a compelling transistor with a subthreshold swing of SS=75~mV/dec and a DIBL of 32~mV at the saturation region operation  (i.e. for drain bias of $V_{d}=0.7$~V). However, the measurements also show that outside of the operating regime, i.e. for negative gate voltages ($V_{g}<0$~V), the drain current ($I_{d}$), instead of the exponential decrease with the gate voltage, first exhibits saturation and then increases as $V_{g}$ reduces. This observed behavior is clearly not thermionic in its origin, as the thermionic current (i.e. over-the-barrier current) must decrease exponentially with the increase of the channel barrier height.

The need to understand the origin of this deviation from thermionic behavior, indicated as leakage current in Fig.~\ref{Fig: ideal GAAFET}, is motivated by two main factors. Firstly, this analysis is crucial for determining the scalability limits and guiding the optimizations required in the future GAAFET designs to further reduce the off-state current $I_{off}$. Secondly, it is important to analyze their performance degradation under elevated radiation environments\textemdash such as those encountered in space (e.g. satellites or visionary AI-computing projects \cite{Google_space}) and in nuclear reactors\textemdash as studies show that, at low to moderate radiation levels, the negative-gate-voltage regime is strongly affected, whereas the normal operating regime remains largely unchanged \cite{Titze:2024}.

In traditional MOSFETs or FinFETs \cite{Chen:1987,Kerber:2013}, the drain leakage at negative gate voltages has been associated with gate-induced drain leakage (GIDL), which is caused by a high electric field near the gate/drain overlap region. This strong field causes pronounced band bending, enabling band-to-band tunneling (BTBT). As a result, an undesired current flows from the grounded substrate to the drain, even when the transistor is supposed to be off \cite{Sze:2006,Chen:1987}. However, in GAAFETs, such a mechanism is unlikely, as the channel-active part of the device is completely isolated electrically from the substrate by a high-k dielectric material \cite{Loubet:2017}. Additionally, another type of GIDL has been reported in ultra-scaled transistors, in which a sufficiently high negative gate voltage can lead to a tunneling current from the gate into the channel valence band, followed by band-to-band tunneling \cite{Kerber:2013}. This tunneling can also be enhanced by the presence of traps/defects near the dielectric material/semiconductor interface or inside the dielectric material. However, in IBM's GAAFET designs\cite{Loubet:2017}, the operating drain biases ($0.05-0.7$~V) have been chosen to minimize this band-to-band tunneling due to the lack of overlap between the conduction and valence bands near the gate-channel interface.

We here propose that in GAAFETs the observed drain leakage at negative gate voltages stems from the gate-to-drain tunneling current, which is further enhanced by the drain-induced \emph{dielectric} barrier lowering. Indeed, our simulations confirm our hypotheses. Simulations are performed using a predictive, first-principles quantum transport approach based on the Non-Equilibrium Green’s Function (NEGF) formalism \cite{Commun_Phys:2021,Mendez_CS:2022,Mendez:2023}. 

In the following, we first analyze the source-drain current vs gate voltage characteristics of an idealized GAAFET, i.e. without gate leakage, using the parameters of the IBM's device presented in Ref.~\cite{Loubet:2017}. Then, we explore the impact of gate leakage on the device and demonstrate that it is the gate-to-drain tunneling current that is responsible for the observed drain leakage at negative gate voltages.

\section{Results and discussion}\label{sec:results and dicussion}

\begin{table}
    \centering
    \begin{tabular}{lc}
     Source/Drain doping [cm$^{-3}$]     & $3.0 \times 10^{20}$ \\
     Channel doping [cm$^{-3}$]          & $5.0 \times 10^{17}$  \\
     Gate length ($L_g$) [nm]                    & 12.0\\
     Metal work function [eV]            & {4.5}\\
     Effective Nanosheet width ($W_{Si}$) [nm]     & 30.0 \\
     Nanosheet thickness ($T_{Si}$) [nm] & 5.0\\
     high-k dielectric thickness [nm]               & 2.0 \\
     high-k dielectric barrier height [eV]          & 1.6 \cite{HfO2:2000,HfO2}   \\
     high-k dielectric constant                    & 20.0 \cite{Lin:2002} \\
    \end{tabular}
    \caption{Material and device parameters used in the simulations. The device parameters are chosen accordingly with the values reported in Ref.~\cite{Loubet:2017}. HfO$_2$ is assumed as the high-k dielectric. Effective nanosheet width is defined as the number of nanosheets times their width.}
    \label{tab:Device paramters}
\end{table}

\begin{figure}[th!]
\includegraphics[width=0.95\columnwidth]{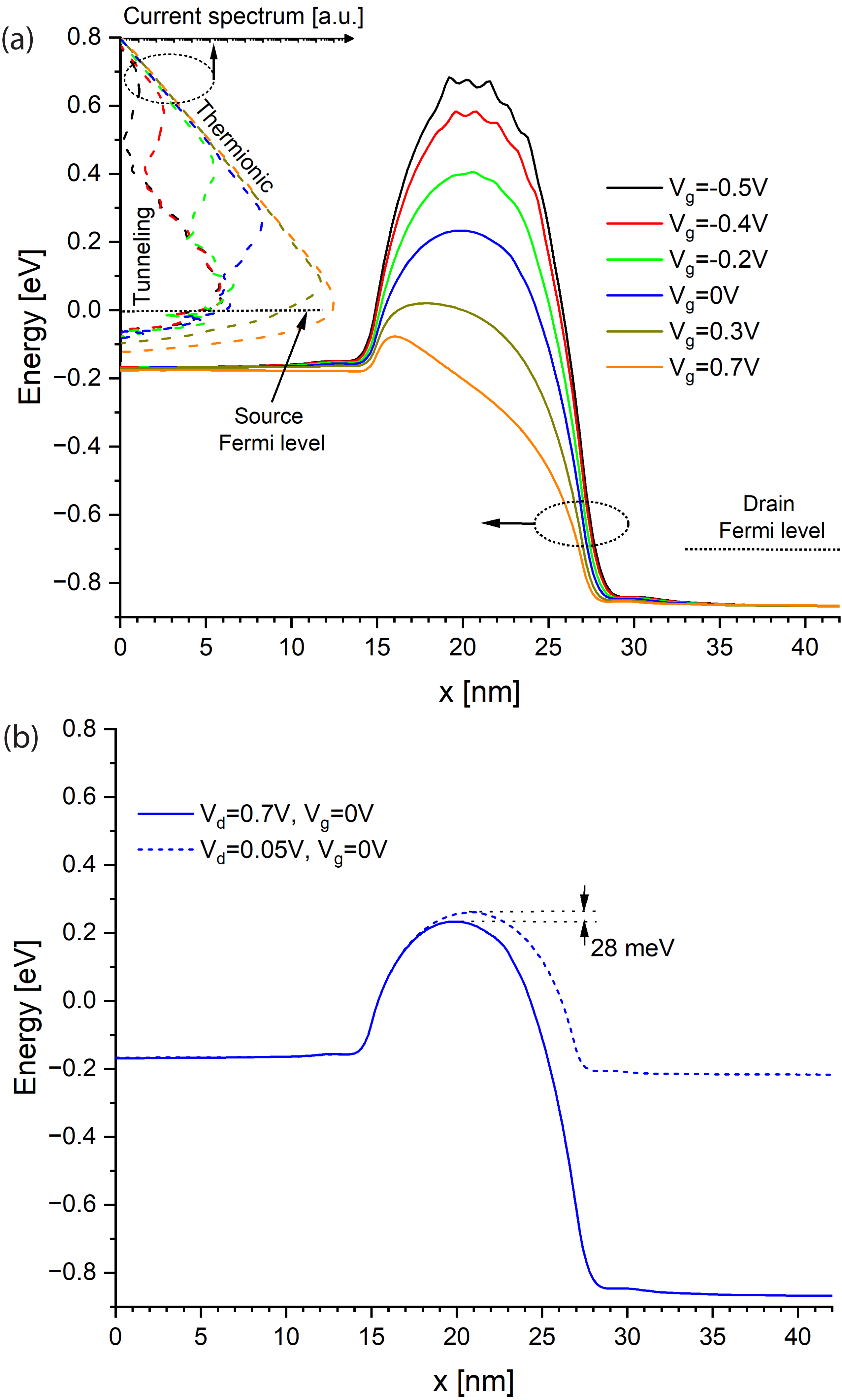}
\caption{(a) Effective potentials along the device channel for several gate voltages, illustrating how the channel effective-potential barrier change with the gate voltage. The applied drain bias is $V_d=0.7$~V.  The inset illustrates the source-drain current spectrum $j_{SD}(E)$, indicating the corresponding thermionic and tunneling contributions. (b) Effective potentials along the channel for two distinct drain biases, $V_d=0.05$ and $0.7$~V, illustrating the DIBL (drain-induced barrier lowering) effect commonly observed in short-channel devices.}
\label{Fig: Dirichlet BC model: current spectrum}
\end{figure}

\begin{figure*}[th!]
\centering
\includegraphics[width=1.0\linewidth]{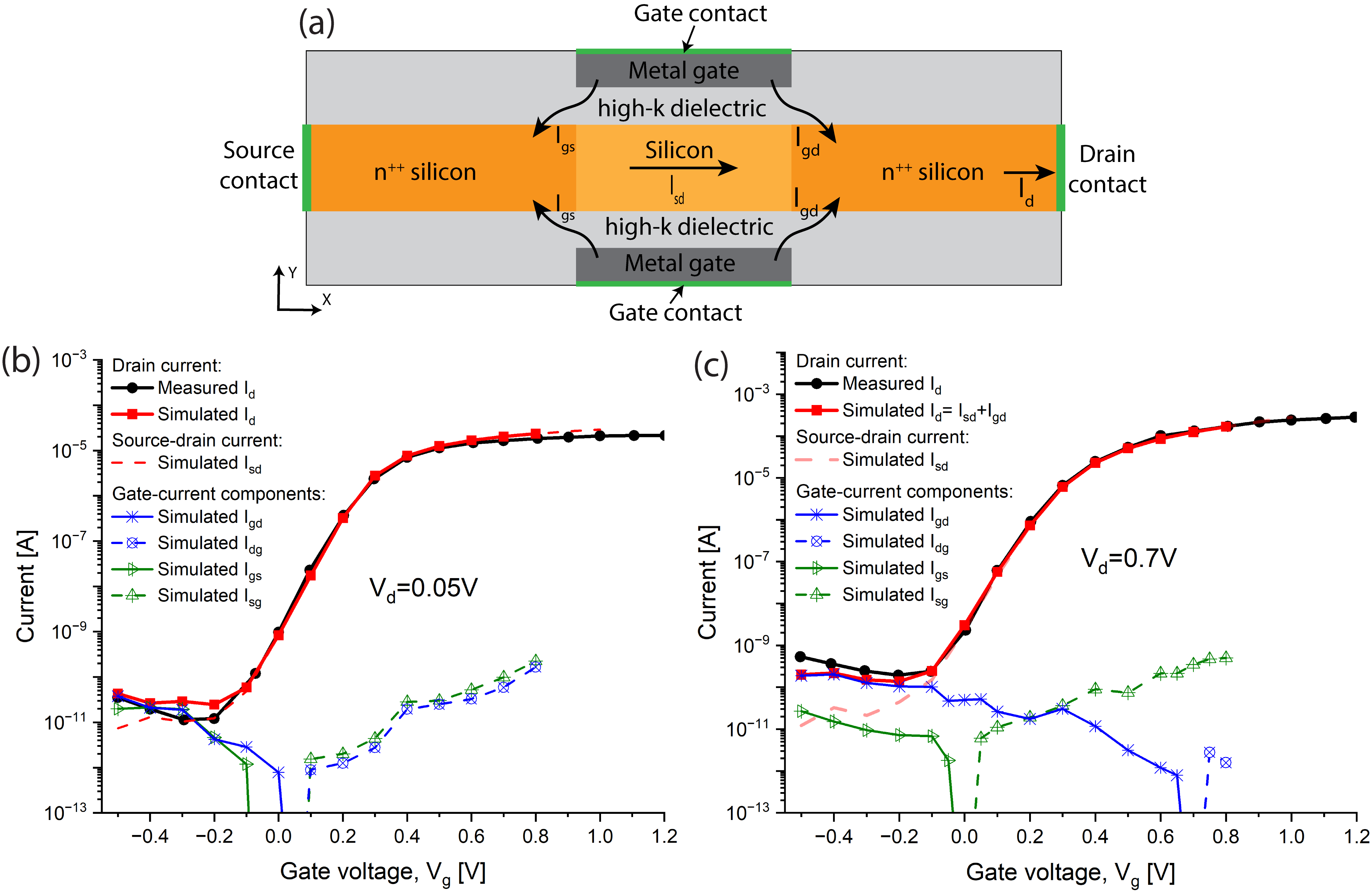}
\caption{(a) Cross-section schematic of the single-sheet GAAFET device used in this work to investigate the gate leakage. $I-V_{g}$ characteristics and the corresponding gate leakage current contributions for $V_d=0.05$~V in (b) and $V_d=0.7$~V in (c). Measured current reprinted with permission from Ref.~\citenum{Loubet:2017}} 
\label{Fig: Full gate: I-V}
\end{figure*}

Fig.~\ref{Fig: ideal GAAFET}(a) shows the simulated characteristics of a single-sheet GAAFET, where gate leakage is assumed to be negligible, as shown in (b). For simulations, we used the device parameters summarized in Table~\ref{tab:Device paramters}. Those were chosen to be consistent with the parameters reported in Ref.~\cite{Loubet:2017}. All simulations presented here are conducted from first-principles in the sense that they do not require parameter adjustments to fit the experimental data. A more detailed description of the used methodology is provided in Appendix~\ref{sec:method}. As the results indicate, for positive gate voltages and both drain biases, the predicted current values agree very well with the IBM's measurements presented in Ref.~\cite{Loubet:2017}.
However, for negative gate voltages, the simulations underestimate the drain current values compared to the experiments, especially for higher drain biases. Additionally, simulations also show the deviation from the ideal thermionic trend at sufficiently negative gate voltages as illustrated in the figure.

Firstly, to elucidate the deviation of the simulated currents from the thermionic trend indicated in Fig.~\ref{Fig: ideal GAAFET}(a), it is necessary to examine the effective potential in the channel and the current spectrum shown in Fig.~\ref{Fig: Dirichlet BC model: current spectrum}~(a). The main panel illustrates how the effective potential changes along the device channel, from the source to the drain, with the gate voltage. The potential barrier increases with the decrease of the gate voltage as electrons are expelled from the channel. The inset to the figure shows the quantum-mechanical current spectrum, i.e. the energy distribution of the electron current. In non-equilibrium quantum mechanics, the current spectrum $j_{sd}(E)$ determines the total current as follows: $I_d=\int_{-\infty}^{\infty}j_{sd}(E) dE$. The current spectrum provides valuable information because, together with the potential profile, it can be used to distinguish the respective contributions of tunneling and thermionic currents. As illustrated in the inset plot, the \emph{thermionic current} corresponds to the portion of the current spectrum that exhibits an exponential dependence on energy. In contrast, the part of the spectrum that is largely independent of the gate voltage corresponds to the  \emph{tunneling current}. Between these two regions, there exists a transitional component associated with a combination of both contributions. For  gate voltages lower than $V_{g}=-0.1~V$, the gate-voltage-independent tunneling component becomes the dominant contribution to the total current, leading to the current saturation observed in the simulations at more negative gate voltages. 

Fig.~\ref{Fig: Dirichlet BC model: current spectrum} (b) shows the effective potential along the device channel, illustrating the drain-induced channel barrier lowering (DIBL) effect that is commonly observed in short channel devices \cite{Sze:2006}. The applied drain bias reduces the  effective barrier in the channel, thus increasing the source-to-drain current. For the off-sate, the simulations indicate a DIBL of approximately $28$~meV, compared to the estimated DIBL of $32$~meV reported in Ref.~\cite{Loubet:2017}

\begin{figure*}[t!]
\centering
\includegraphics[width=0.7\linewidth]{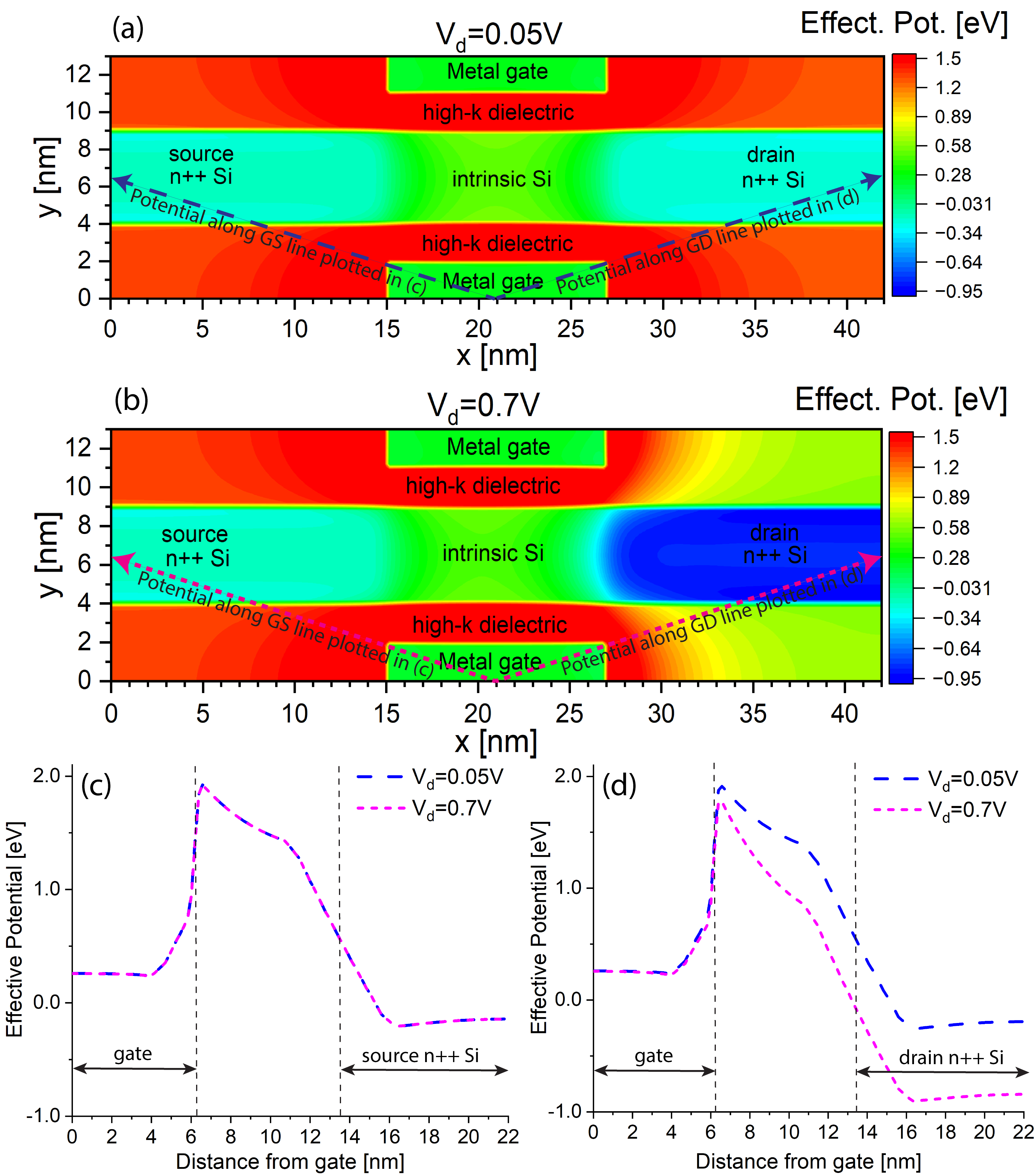}
\caption{Contour-color maps of the effective potential for $V_d=0.05$~V in (a) and for $V_d=0.7$~V in (b). (c) Effective potential along the line drawn in (a,b) between gate and source for a $V_d$ of 0.05 and 0.7~V.  (d) Effective potential along the line drawn in (a,b) between gate and drain for a $V_d$ of 0.05 and 0.7~V. The gate voltage for all cases (a-d) is $V_g=-0.3$~V.}
\label{Fig: Elec potential}
\end{figure*}

As shown in the above analyses of an idealized GAAFET without gate leakage, source-drain tunneling current alone cannot account for the observed behavior for negative voltages. We next analyze, for the first time, the contribution of the gate leakage in GAAFETs from the first principles as illustrated in Fig.~\ref{Fig: Full gate: I-V} (a). 
Figs.~\ref{Fig: Full gate: I-V} (b) and (c) present the simulated total drain current $I_d$, along with the current components $I_{sd}$, $I_{gs}$ and $I_{gd}$, which correspond to the source-drain, gate-source and gate-drain currents, respectively, as functions of gate voltages for drain biases of $V_d=0.05$ and $0.7$~V. Here, note that we use the convention for the sign of current in quantum mechanics, determined by the probability current density. First, the simulations show overall excellent qualitative and near-quantitative agreement with the measured drain currents at both drain biases, confirming that the previously discussed discrepancies arise from gate-drain leakage. For positive gate voltages, although the simulations reveal the existence of gate leakage, the gate tunneling current is several orders of magnitude smaller than the dominant source-drain current. The situation changes markedly for negative gate voltages: the gate–drain current becomes significant, even exceeding the source–drain current and becoming the main contribution to the total drain current, especially at larger applied drain biases. Furthermore, our simulations also accurately reproduce the influence of the drain bias on the gate–drain current observed experimentally under negative gate voltage, indicating that the drain bias induces an effective \emph{dielectric} barrier lowering between the gate and drain. This drain-induced barrier lowering is analogous to the conventional DIBL effect observed in short-channel devices between the source and drain, and exists in GAAFETs, as shown in Fig.~\ref{Fig: Dirichlet BC model: current spectrum} (b). However, in this case, the tunnel barrier affected by the drain bias is the dielectric barrier between the gate and drain, and the effect is even more pronounced than the DIBL observed in the channel.

We next examine in detail how the drain bias affects the gate-drain tunneling current. Fig.~\ref{Fig: Elec potential} shows the effective potentials across the GAAFET for $V_d=0.05$ and $0.7$~V in (a) and (b), respectively. The figure also include the 1D-effective potentials in (c) and (d) that represent the barrier between the gate-source and gate-drain, respectively. From these results, it can be observed that as the drain bias increases from 0.05~V to 0.7~V, the effective barrier between the gate and drain (dielectric barrier) decreases considerably due to their close proximity, resulting in a drain-induced barrier lowering effect. Consequently, the reduced effective barrier between the metal gate and the highly doped drain leads to the significant increase of the gate-drain tunneling leakage at negative gate voltages. In contrast, the barrier between the gate and the source remains largely unaffected (Fig.~\ref{Fig: Elec potential}~(c)) that results in very similar contributions of the gate-to-source leakage currents at different drain biases (Figs.~\ref{Fig: Full gate: I-V}~(b),(c), green curves). 

It can be also observed from Fig.~\ref{Fig: Full gate: I-V} (c)  that the simulated drain currents slightly underestimate the measured ones for negative gate voltages at a drain bias of $0.7$~V. This small discrepancy can be attributed to the fact that valence electrons are not included in the calculations. Consequently, electron tunneling through the dielectric material into the semiconductor valence band, followed by band-to-band tunneling from the valence to the conduction band near the gate region, is not captured in these results. 
The good agreement with existing measurements indicates that the main current mechanism between gate and drain is the one reported in this work, the direct tunneling, i.e. electrons tunneling directly into the conduction band. This underestimation of the drain current is not observed at a drain bias of $V_d = 0.05$~V (Fig.~\ref{Fig: Full gate: I-V} (b)), as this voltage is insufficient to lower the conduction band below the valence band near the metal gate–silicon channel interface, thus preventing band-to-band tunneling mechanism.

\section{Conclusions}\label{sec:conclusions}

GAAFETs have been demonstrated to be strong candidates to replace FinFETs for continued semiconductor scaling. However, electrical measurements reveal an intriguing drain current increase at negative gate voltages. This behavior of the drain current at negative gate voltages deviates from the ideal thermionic trend expected. In MOSFETs, this significant drain leakage for negative gate voltages has been associated with GIDL. However, in GAAFETs, such mechanism is unlikely, as the channel-active part of the device is completely isolated electrically from the substrate by a high-k dielectric material. Through predictive first principle simulations, it has been demonstrated that the current increase with increasing negative gate voltages stems from the gate-to-drain leakage tunneling current. Additionally, it has been found that the observed gate leakage is enhanced by a drain-induced barrier lowering between the metal gate and the highly doped drain across the dielectric material region: an increase in the drain bias (e.g., from 0.05 V to 0.7 V) causes the effective dielectric barrier reduction due to their close proximity. This drain-induced barrier lowering effect facilitates the gate-drain tunneling leakage at negative gate voltages.

The deviation from the thermionic behavior at the deep sub-threshold regime (negative gate voltages) due to gate-to-drain leakage currents has important consequences for future device designs. The results presented in this work indicate that to continue the gate-all-around transistor scaling to physical gate lengths of sub-$10$~nm, one has to solve at least two principle problems: 1) the reduction of the gate leakage current induced by the drain bias and 2) the reduction of the source-to-drain tunneling current. The first problem can be readily mitigated, for instance, with designs that reduce the gate-to-drain tunneling by increasing the high-k dielectric thickness specifically at the drain side (i.e. without worsening the control of the gate over the channel) or replacing the drain side dielectric with silicon dioxide, which has a high tunnel barrier of about 3~eV \cite{HfO2}. The source-to-drain tunneling current presents a more serious difficulty for further scaling, as it would increase exponentially with the physical gate length reduction. It is therefore likely that new material designs, confinement-based band structure engineering, and other novel switching techniques are necessary for GAAFETs of sub-$10$~nm gate lengths.

Finally, our studies and conclusions are based on the GAAFET design reported in Ref.~\cite{Loubet:2017} in 2017. Since then, the industry may have introduced improvements to mitigate some or all of the mentioned challenges; however, such advancements have not yet been publicly disclosed.

\begin{acknowledgments}
This work was supported in part by the U.S. the Department of Energy’s Advanced Simulation and Computing (ASC) Program and the Laboratory Directed Research and Development (LDRD) program. 
Juan P. Mendez and  Coleman Cariker contributed equally to this work.
Sandia National Laboratories is a multimission laboratory managed and operated by National Technology and Engineering Solutions of Sandia, LLC., a wholly owned subsidiary of Honeywell International, Inc., for the U.S. Department of Energy’s National Nuclear Security Administration under contract DE-NA-0003525. This paper describes objective technical results and analysis. Any subjective views or opinions that might be expressed in the paper do not necessarily represent the views of the U.S. Department of Energy or the United States Government. This article has been authored by an employee of National Technology \& Engineering Solutions of Sandia, LLC under Contract No. DE-NA0003525 with the U.S. Department of Energy (DOE). The employee owns all right, title and interest in and to the article and is solely responsible for its contents. The United States Government retains and the publisher, by accepting the article for publication, acknowledges that the United States Government retains a non-exclusive, paid-up, irrevocable, world-wide license to publish or reproduce the published form of this article or allow others to do so, for United States Government purposes. The DOE will provide public access to these results of federally sponsored research in accordance with the DOE Public Access Plan https://www.energy.gov/downloads/doe-public-access-plan. 
\end{acknowledgments}

\section*{Data Availability}
The data that support the findings of this study are available from the corresponding author upon reasonable request.




\appendix

\section{Methodology}\label{sec:method}

The simulations in this work are conducted using the open-system charge self-consistent Non-Equilibrium Green Function (NEGF) Keldysh formalism \cite{Keldysh:1965,Datta:1997}, implemented using the Contact Block Reduction (CBR) method \cite{Mamaluy:2003,Mamaluy_2004,Sabathil_2004,Mamaluy:2005,Khan:2007,Khan_2008,Gao:2014,Mendez:2021}. The CBR method allows a very efficient calculation of the density matrix, transmission function, etc. of an arbitrarily shaped, multiterminal two- or three-dimensional open device within the NEGF formalism and scales linearly $O(N)$ with the system size $N$.  We note that the applicability of the CBR method to multi-terminal systems enables the fully quantum-mechanical treatment for all three current-carrying contacts (the source, drain and gate) , which significantly increases the accuracy of the gate leakage calculation.

Within this framework, we solve charge self-consistently the open-system Schr\"{o}dinger equation and the non-linear Poisson equation \cite{Mamaluy:2003,Mamaluy:2005,Gao:2014}. Here we employ a multi-band effective mass tensor approximation for the kinetic energy operator with three non-equivalent double-degenerated silicon valleys defined by the longitudinal effective mass along each of the Cartesian axes and the transverse effective masses along the other axes. For the charge self-consistent solution of the non-linear Poisson equation, we use a combination of the predictor-corrector approach and Anderson mixing scheme \cite{Khan:2007,Gao:2014}. First, an eigenvalue problem is solved, for each valley, for a specially defined closed system, while taking into account the electrostatic Hartree potential $\phi^H(\textbf{r}_i)$ and the exchange and correlation potential $\phi^{XC}(\textbf{r}_i)$ \cite{PerdewZunger:1981}. Second, the local density of states (LDOS) of the open system, $\rho(\textbf{r}_i,E)$, and the electron density, $n(\textbf{r}_i)$, are computed using the CBR method for each iteration. Then, the electrostatic potential and the carrier density are used to calculate the residuum $F$ of the Poisson equation
\begin{equation}
\big|\big|\textbf{F}[\boldsymbol\phi^H(\textbf{r}_i)]\big|\big|=\big|\big|\textbf{A}\boldsymbol\phi^H(\textbf{r}_i) - (\textbf{n}(\textbf{r}_i)-\textbf{N}_D(\textbf{r}_i)+\textbf{N}_A(\textbf{r}_i))\big|\big|,
\end{equation}
where $\textbf{A}$ is the matrix derived from the discretization of the Poisson equation, and $\textbf{N}_D$ and $\textbf{N}_A$ are the total donor and acceptor doping densities arrays, respectively. If the residuum is larger than a predetermined threshold $\epsilon$, the Hartree potential is updated using the predictor-corrector method, together with the Anderson mixing scheme. Using the updated Hartree potential and the corresponding carrier density, the exchange-correlation is computed again for the next step, and an iteration of the Schr\"{o}dinger-Poisson equations is repeated until the convergence is achieved with $\big|\big|\textbf{F}[\boldsymbol\phi^H(\textbf{r}_i)]\big|\big|<\epsilon=1\times 10^{-6}$~eV.
In our simulations we employed the standard values of the inertial effective mass tensor for electrons, $m_l = 0.98 \times m_e$, $m_t = 0.19 \times m_e$, and the dielectric constant of silicon $\epsilon_{Si}=11.7$.  The simulations and measurements are carried out at 300~K.  For ineslastic scattering, we have employed the model presented in Ref.~\cite{mamaluy:2024}.

The effective potential presented in this work (Figs.~\ref{Fig: Dirichlet BC model: current spectrum} and \ref{Fig: Elec potential}) is the sum of the Hartree (electrostatic) potential, the exchange and correlation corrections within the local density approximation (LDA), and the band offsets due to the change in the work-function values for different materials (e.g. silicon/dielectric).

\bibliography{reference}

\end{document}